\documentclass{article}
\usepackage{amsmath,amssymb}
\usepackage{mathrsfs}
\usepackage[dvips]{graphicx}
\newenvironment{ol}[1]{
\begin{enumerate}
\setlength{\partopsep}{#1pt}
\setlength{\topsep}{#1pt}
\setlength{\itemsep}{#1pt}
\setlength{\parsep}{#1pt}
\setlength{\parskip}{#1pt}}{
\end{enumerate}
}

\title{A model for learning to segment temporal sequences, utilizing a mixture of RNN experts together with adaptive variance}

\author{
Jun Namikawa \\
{\small Brain Science Institute, RIKEN}\\
{\small 2-1 Hirosawa, Wako-shi, Saitama, 351-0198 Japan}\\
{\small E-mail \texttt{jnamika@bdc.brain.riken.go.jp}}\\
\and
Jun Tani \\
{\small Brain Science Institute, RIKEN}\\
{\small 2-1 Hirosawa, Wako-shi, Saitama, 351-0198 Japan}\\
{\small E-mail \texttt{tani@brain.riken.go.jp}}\\
}

\date{}

\begin{document}

\maketitle
\begin{flushleft}
{\bf Acknowledgment}\\
Requests for reprints should be sent to Jun Namikawa, Brain Science
 Institute, RIKEN. 2-1 Hirosawa, Wako-shi,
 Saitama, 351-0198 Japan \\
{\small Tel +81-48-467-6467, FAX +81-48-467-7248}
\end{flushleft}
\begin{flushleft}
{\bf Key Words}\\
recurrent neural network, mixture of experts,
maximum likelihood estimation, self-organization, segmentation of temporal sequences
\end{flushleft}

\begin{abstract}
This paper proposes a novel learning method for a mixture of recurrent neural network (RNN) experts model, which can acquire the ability to generate desired sequences by dynamically switching between experts.
Our method is based on maximum likelihood estimation, using a gradient descent algorithm.
This approach is similar to that used in conventional methods; however, we modify the likelihood function by adding a mechanism to alter the variance for each expert.
The proposed method is demonstrated to successfully learn Markov chain switching among a set of $9$ Lissajous curves, for which the conventional method fails.
The learning performance, analyzed in terms of the generalization capability, of the proposed method is also shown to be superior to that of the conventional method.
With the addition of a gating network, the proposed method is successfully applied to the learning of sensory-motor flows for a small humanoid robot as a realistic problem of time series prediction and generation.
\end{abstract}


\section{Introduction} \label{section:introduction}

How to learn complex temporal sequence patterns by means of neural network models has been a challenging problem.
Recurrent neural networks (RNNs) \cite{Jordan86,Elman90,Pollack91,Zipser89} have been one of the most popular models applied to temporal sequence learning.
RNNs can learn sensory-motor sequence patterns \cite{Jordan86,Jordan88a}, symbolic sequences with grammar \cite{Elman90,Pollack91} and continuous spatio-temporal patterns \cite{Zipser89}.

In spite of the considerable amount of RNN research carried out since the mid 1980s, it has been thought that RNNs cannot be scaled to be capable of learning complex sequence patterns, especially when the sequence patterns to be learnt contain long-term dependency \cite{Bengio94}.
This is due to the fact that the error signal cannot be propagated effectively in long-time windows of sequences using the back-propagation through time (BPTT) algorithm \cite{Rumelhart86}, because of the potential nonlinearity of the RNN dynamics \cite{Bengio94}.

There have been some breakthroughs in approaches to this problem.
Hochreiter and Schmidhuber \cite{Hochreiter97,Schmidhuber2002} proposed a method called ``Long Short-Term Memory'' (LSTM). 
LSTM learns to bridge minimal time lags in long time steps by enforcing constant error flow through ``constant error carousels'' within special units. 
These units learn to open and close access to the constant error flow.
Echo state networks \cite{Jaeger2001,Jaeger2004} and a very similar approach, liquid state machines \cite{Maass2002}, have recently attracted significant attention.
Echo state networks possesses a large pool of hidden neurons with fixed random weights, and the pool is capable of rich dynamics.
Jaeger \cite{Jaeger2004} demonstrated that an echo state network can successfully learn the Mackey-Glass chaotic time series which is a well-known benchmark system for time series prediction.

Tani and Nolfi \cite{Tani99} investigated the same problems, but from a different angle, focusing on the idea of compositionality for sensory-motor learning.
The term compositionality was adopted from the ``Principle of Compositionality'' \cite{Evans81} in linguistics, which claims that the meaning of a complex expression is determined by the meanings of its constituent
expressions and the rules used to combine them.
This principle, when translated to the sensory-motor learning context, leads to the assertion that diverse and complex patterns can be learned to be generated by adaptively combining a set of re-usable behavior primitives.
Here, acquiring behavior primitives requires a mechanism for autonomously segmenting a continuously experienced sensory-motor flow into reusable chunks.
Tani and Nolfi (1998) proposed a scheme for hierarchical segmentation of the sensory-motor flow, applying the idea of a mixture of experts \cite{Jacobs91,Jordan94,Wolpert98} to hierarchically organized RNNs.
Consider a network consisting of multiple local RNNs, organized in levels. At the base level, each RNN competitively learns to be an expert at predicting/generating a specific sensory-motor profile.
As the winner among the expert modules changes, corresponding to structural changes in the sensory-motor flow, it is considered that the sensory-motor flow is segmented, by switching between winners.
Meanwhile, the RNN experts at the higher level learn the winner switching sequence patterns with much slower time constants compared to those at the sensory-motor level.
The higher level learns not for details of sensory-motor patterns but for abstraction of primitive sequences with long-term dependency.

Although the scheme of the mixture of RNN experts seems to be tractable, in reality there are potential problems arising from scaling \cite{TaniSMC2007}.
It is known that if the number of modules increases, segmenting the sequence becomes unstable and learning by a gating mechanism tends to fail.
This is due to near-miss problems in matching the current sensory-motor pattern to the best local RNN among others, which have acquired similar pattern profiles.
Since the number of modules determines the scalability of the system, it is desirable that learning proceeds stably with larger numbers of modules.

In the present study, a novel learning method is presented for a mixture of RNN experts model.
The proposed learning method allows a sequence to be segmented into reusable blocks without loss of stability with increasing number of learning modules.
A maximum likelihood estimation based on the gradient descent algorithm is employed, similar to the conventional learning method for mixture of RNN experts \cite{Tani99,IgariNC2006}, although the present likelihood function differs from those presented previously.
In the proposed method, the weighting factor for the allocation of blocks to modules is changed adaptively via a variance parameter for each module.
The superior performance of the proposed method is demonstrated through application to a problem that involves learning to extract a set of reusable primitives from data constructed by stochastically combining multiple Lissajous curve patterns.
The proposed learning method is compared with the conventional method on the basis of generalization capability as a measure of the extent to which compositionality can be achieved in modular networks.
Finally, the proposed method is applied in combination with a gating network to learn the sensory-motor flows for a small humanoid robot as a realistic problem.

\section{Model} \label{section:model}

In this section, we define a model known as a mixture of recurrent neural network (RNN) experts model.
A mixture of RNN experts is simply a mixture of experts model for which the learning modules are RNNs.
The model equations are defined as follows:
\begin{equation} \label{equation:mixture_of_rnn_experts1}
\boldsymbol{u}_{n}^{(i)} = \big(1-\epsilon\big)\boldsymbol{u}_{n-1}^{(i)} + \epsilon\big( W_1^{(i)} \boldsymbol{x}_n + W_2^{(i)} \boldsymbol{c}_{n-1}^{(i)} + \boldsymbol{v}_1^{(i)}\big),
\end{equation}
\begin{equation} \label{equation:mixture_of_rnn_experts2}
\boldsymbol{c}_n^{(i)} = \tanh(\boldsymbol{u}_{n}^{(i)}),
\end{equation}
\begin{equation} \label{equation:mixture_of_rnn_experts3}
\boldsymbol{y}_n^{(i)} = \tanh(W_3^{(i)} \boldsymbol{c}_{n}^{(i)} + \boldsymbol{v}_2^{(i)}),
\end{equation}
\begin{equation} \label{equation:mixture_of_rnn_experts4}
\boldsymbol{y}_n = \sum_{i=1}^{N}g_n^{(i)} \boldsymbol{y}_n^{(i)},
\end{equation}
where $N$ is the number of expert modules, $\epsilon$ is a time constant satisfying $0 \leq \epsilon \leq 1$, $\boldsymbol{x}_n$ and $\boldsymbol{y}_n$ are the input and output vectors of the model at time $n$, respectively.
For each $i$, $\boldsymbol{u}_n^{(i)}$, $\boldsymbol{c}_n^{(i)}$ and $\boldsymbol{y}_n^{(i)}$ denote internal potential of context neurons, states of context neurons and states of output neurons of the module $i$, respectively.
The matrices $W_1^{(i)}, W_2^{(i)}, W_3^{(i)}$ and vectors $\boldsymbol{v}_1^{(i)}, \boldsymbol{v}_2^{(i)}$ are parameters of the module $i$.
A gate opening value for the module $i$ is denoted by $g_n^{(i)}$, and it is assumed that $g_n^{(i)} \geq 0$ and $\sum_{i=1}^{N}g_n^{(i)} = 1$.
The gate opening vector $\boldsymbol{g}_n$ represents the winner-take-all competition among modules to determine the output $\boldsymbol{y}_n$.
The determination of the gate opening vector $\boldsymbol{g}_n$ is explained in section \ref{subsection:learning_rule}.

\subsection{Learning method} \label{subsection:learning_rule}

The proposed learning method is defined on the basis of the probability distribution of the mixture of RNN experts.
Here the parameters of learning module $i$ are denoted by $\vartheta_i = \big(W_1^{(i)}, W_2^{(i)},W_3^{(i)}, \boldsymbol{v}_1^{(i)}, \boldsymbol{v}_2^{(i)}, \boldsymbol{u}_0^{(i)}\big)$, and a gate opening vector $\boldsymbol{g}_n$ is described by a variable $\boldsymbol{\beta}_n$ such as
\begin{equation} \label{equation:gate_opening_values}
g_n^{(i)} = \frac{\exp(\beta_n^{(i)})}{\sum_{k=1}^N \exp(\beta_n^{(k)})}.
\end{equation}

Let $X = (\boldsymbol{x}_n)_{n=1}^T$ be an input sequence, $\boldsymbol{\beta}_n$ and $\boldsymbol{\gamma}$ be parameters, where $\boldsymbol{\gamma}$ is a set of parameters given by $\gamma_i = (\vartheta_i, \sigma_i)$.
Given $X$, $\boldsymbol{\beta}_n$, and $\boldsymbol{\gamma}$, the probability density function (p.d.f.) of output $\boldsymbol{y}_n$ is defined by
\begin{equation}
p(\boldsymbol{y}_n ~|~ X, \boldsymbol{\beta}_n, \boldsymbol{\gamma}) = \sum_{i=1}^{N} g_n^{(i)} p(\boldsymbol{y}_n ~|~ X, \gamma_i),
\end{equation}
where $p(\boldsymbol{y}_n | X, \gamma_i)$ is given by
\begin{equation} \label{equation:mixture_of_normal_distribution}
p(\boldsymbol{y}_n | X, \gamma_i) = \Big(\frac{1}{\sqrt{2\pi}\sigma_i}\Big)^{d}\exp(-\frac{||\boldsymbol{y}_{n}^{(i)} - \boldsymbol{y}_n||^2}{2\sigma_i^2}),
\end{equation}
$d$ is the output dimension, and $\boldsymbol{y}_{n}^{(i)}$ is an output of the module $i$ computed by the equations (\ref{equation:mixture_of_rnn_experts1}), (\ref{equation:mixture_of_rnn_experts2}) and (\ref{equation:mixture_of_rnn_experts3}) with parameter $\vartheta_i$.
Thus, the output of the model is governed by a mixture of normal distributions.
This equation results from the assumption that an observable sequence data is embedded in additive Gaussian noise.
It is well known that minimizing the mean square error is equivalent to maximizing the likelihood determined by a normal distribution for learning in a single neural network.
Therefore, equation (\ref{equation:mixture_of_normal_distribution}) is a natural extension of a neural network for learning.
The details of the derivation of equation (\ref{equation:mixture_of_normal_distribution}) is given in \cite{Jacobs91}.

Given a parameter set $\boldsymbol{\theta} = \big((\boldsymbol{\beta}_n)_{n=1}^{T}, \boldsymbol{\gamma}\big)$ and an input sequence $X$, the probability of an output sequence $Y = (\boldsymbol{y}_n)_{n=1}^T$ is given by
\begin{equation}
p(Y ~|~ X, \boldsymbol{\theta}) = \prod_{n=1}^{T} p(\boldsymbol{y}_n ~|~ X, \boldsymbol{\beta}_n, \boldsymbol{\gamma}).
\end{equation}
The likelihood function $L$ of the data set $D = (X, Y)$, parametrized by  $\boldsymbol{\theta}$, is denoted by 
\begin{equation}
L(D, \boldsymbol{\theta}) = p(Y ~|~ X, \boldsymbol{\theta}) \varphi(\boldsymbol{\theta}),
\end{equation}
where $\varphi(\boldsymbol{\theta})$ is the p.d.f. of the prior distribution given by
\begin{equation} \label{equation:prior_distribution}
\varphi(\boldsymbol{\theta}) = \prod_{n=1}^{T-1} \prod_{i=1}^{N} \frac{1}{\sqrt{2\pi}\varsigma}\exp(-\frac{(\beta_{n+1}^{(i)} - \beta_n^{(i)})^2}{2\varsigma^2}).
\end{equation}
This equation means that the vector $\boldsymbol{\beta}_n$ is governed by $N$-dimensional Brownian motion.
The prior distribution has the effect of suppressing the change of gate opening values.

The learning method proposed here is to choose the best parameter $\boldsymbol{\theta}$ by maximizing (or integrating over) the likelihood $L(D, \boldsymbol{\theta})$ with training data $D$.
Concretely, we use the gradient descent method with a momentum term as the training procedure.
The update rule for the model parameter is 
\begin{equation} \label{equation:gradient_descent_method_with_momentum1}
\Delta\theta(t) = \alpha\frac{\partial \ln L(D, \boldsymbol{\theta}(t))}{\partial \boldsymbol{\theta}(t)} + \eta\Delta\theta(t-1),
\end{equation}
\begin{equation} \label{equation:gradient_descent_method_with_momentum2}
\theta(t+1) = \theta(t) + \Delta\theta(t),
\end{equation}
where $\theta(t)$ is the parameter at learning step $t$, $\alpha$ is the learning rate, and $\eta$ is the momentum term parameter.
For each parameter, the partial differential equation $\frac{\partial \ln L(D, \boldsymbol{\theta})}{\partial \boldsymbol{\theta}}$ are given by
\begin{equation} \label{equation:gradient_beta}
\frac{\partial \ln L(D, \boldsymbol{\theta})}{\partial \beta_{n}^{(i)}} = \frac{g_{n}^{(i)} (p(\boldsymbol{y}_n ~|~ X, \gamma_i) -  p(\boldsymbol{y}_n ~|~ X, \boldsymbol{\beta}_n, \boldsymbol{\gamma}))}{p(\boldsymbol{y}_n ~|~ X, \boldsymbol{\beta}_n, \boldsymbol{\gamma})} + \frac{G_{n}^{(i)}}{\varsigma^2},
\end{equation}
\begin{equation}
G_{n}^{(i)} = \left\{ \begin{array}{cl}
\beta_{n+1}^{(i)} - \beta_{n}^{(i)} & \mbox{if $n = 1$}, \\
-\beta_{n}^{(i)} + \beta_{n-1}^{(i)} & \mbox{if $n = T$}, \\
\beta_{n+1}^{(i)} - 2\beta_{n}^{(i)} + \beta_{n-1}^{(i)} & \mbox{otherwise},
\end{array} \right.
\end{equation}
\begin{equation} \label{equation:gradient_vartheta}
\frac{\partial \ln L(D, \boldsymbol{\theta})}{\partial \vartheta_i} = \sum_{n=1}^{T} \frac{g_{n}^{(i)} p(\boldsymbol{y}_n ~|~ X, \gamma_i)}{p(\boldsymbol{y}_n ~|~ X, \boldsymbol{\beta}_n, \boldsymbol{\gamma})}\frac{-1}{2\sigma_i^2}\frac{\partial}{\partial \vartheta_i}||\boldsymbol{y}_{n}^{(i)} - \boldsymbol{y}_n||^2,
\end{equation}
\begin{equation} \label{equation:gradient_sigma}
\frac{\partial \ln L(D, \boldsymbol{\theta})}{\partial \sigma_i} = \sum_{n=1}^{T} \frac{g_{n}^{(i)} p(\boldsymbol{y}_n ~|~ X, \gamma_i)}{p(\boldsymbol{y}_n ~|~ X, \boldsymbol{\beta}_n, \boldsymbol{\gamma})} \Big[ -\frac{d}{\sigma_i} + \frac{||\boldsymbol{y}_{n}^{(i)} - \boldsymbol{y}_n||^2}{\sigma_i^3} \Big].
\end{equation}
Notice that $\frac{\partial}{\partial \vartheta_i}||\boldsymbol{y}_{n}^{(i)} - \boldsymbol{y}_n||^2$ can be solved by the back propagation through time (BPTT) method \cite{Rumelhart86}.
In this paper, we assume that the infimum $\bar{\sigma}$ of the parameter $\sigma_i$ is greater than $0$, because if $\sigma_i$ converges to $0$, then $\Delta\theta(t)$ diverges.

The explanation above refers to a training data set $D$ consisting of a single sequence.
However, the method can be readily extended to the learning of several sequences by calculating the sum of gradients for each sequence.
When several sequences are used as training data, the initial states $\boldsymbol{u}_0^{(i)}$ and the parameters $\boldsymbol{\beta}_n$ of the gate opening vector must be provided for each sequence.

An important difference between the proposed method and the conventional method used in \cite{Tani99,IgariNC2006} is the use of an optimized variance $\boldsymbol{\sigma}$ of the normal distribution.
Indeed, if $\boldsymbol{\sigma}$ is a constant, e.g., $\boldsymbol{\sigma} = \boldsymbol{1}$, then the proposed method is equivalent to the conventional one.
Adapting the variance $\boldsymbol{\sigma}$ results in different learning speeds for different learning modules, because the learning speed of a module $i$ depends on $\sigma_i$, from equation (\ref{equation:gradient_vartheta}).
Moreover, from equation (\ref{equation:gradient_sigma}), $\sigma_i$ decreases if the mean square error of the module $i$ is less than $d\sigma_i^2$.
This fact implies that there exists positive feedback in the learning speed of modules.
When the same sequence blocks are allocated to a set of modules, the learning speed of the best matching module becomes faster than for the other modules in the set due to this feedback mechanism, reinforcing the matching between the module and the allocated blocks.
Hence, this feedback leads to a decision on the winner of the competition among modules.
It is shown later that the adaptive optimization of $\boldsymbol{\sigma}$ plays an important role in segmenting the sequence $D$ into blocks via the gate opening vector $\boldsymbol{g}_n$.

\subsection{Feedback loop with time delay} \label{subsection:open_and_closed_loop}

Let us consider the case in which there exists a feedback loop from output to input with a time delay $\tau$, that is to say, the model is an autonomous system.
In this case, a training data set $D = (X, Y)$ has to satisfy $\boldsymbol{y}_n = \boldsymbol{x}_{n+\tau}$.
At the end of learning, if every output of the model is completely equal to the training data, the model can generate the sequence $Y$ using a feedback loop instead of using the training data $X$ as external input.
In this paper, open-loop dynamics refers to the case in which external inputs are given by training data, and closed-loop dynamics refers to the case involving self-feedback.
In section \ref{section:simulation}, we consider the case in which the model is an autonomous system.

\section{Numerical simulation} \label{section:simulation}

\subsection{Learning} \label{subsection:learning_example}

\begin{figure}
\begin{center}
\scalebox{1.0}{\includegraphics{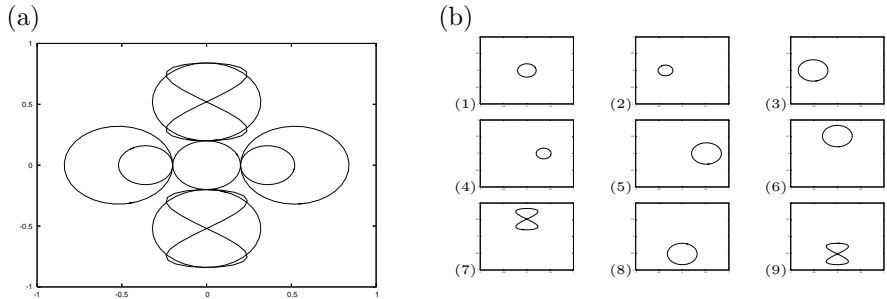}}
\caption{(a) Training data generated by Markov chain switching of $9$ Lissajous curves.
(b) Each Lissajous curve.
The subscript of each figure denotes the index of each Lissajous curve.
The transitions among curves are consonant with continuity of the orbit.}
\label{figure:teaching_data}
\end{center}
\end{figure}

The learning ability of the proposed learning method with optimizing $\boldsymbol{\sigma}$ is compared here that of the conventional method (constant $\boldsymbol{\sigma}$) and a standard RNN model with BPTT \cite{Rumelhart86} as a benchmark.
The architecture of the RNN with BPTT method is the same as that for a module of the mixture of RNN experts model.
The training data is a two-dimensional sequence generated by Markov chain switching of $9$ Lissajous curves, each of period $32$ (see Figure \ref{figure:teaching_data}).
Assume that the transition probabilities of the Markov chain are such that transitions among curves are consonant with orbit continuity.
Since we consider the model which has a feedback loop with time delay $\tau$, the training data set $D = (X, Y)$ satisfies $\boldsymbol{y}_n = \boldsymbol{x}_{n+\tau}$.
We provides details of the training data in Appendix \ref{section:appendix_A}.

We describe the experimental conditions in this section.
The training data was of length $T = 10,000$, and learning was conducted for $300,000$ steps in each model.
The parameter settings were set to $N = 24$ learning modules, $\dim = 10$ context neurons in each learning module, a time constant of $\epsilon = 0.1$, a time delay of $\tau = 5$ for the feedback loop, an infimum of $\bar{\sigma} = 0.05$ for the variance of the normal distribution, a standard deviation of $\varsigma = 1$ for the prior distribution, a momentum of $\eta = 0.9$, and a learning rate of $\alpha = 0.01 / Td$, where $T$ is length and $d$ is the dimension of the training data.
The learnable parameters of the mixture of RNN experts were initialized as $\boldsymbol{\beta}_n = \boldsymbol{0}$ and $\boldsymbol{\sigma} = \boldsymbol{1}$, every element of the matrices $W_1^{(i)}, W_2^{(i)},W_3^{(i)}$ and the vectors $\boldsymbol{v}_1^{(i)}, \boldsymbol{v}_2^{(i)}$ were initialized randomly from the uniform distribution on the interval $(-0.1, 0.1)$, and an initial state $\boldsymbol{u}_0^{(i)}$ for each $i$ was initialized randomly from the interval $[-1,1]$.
In the RNN with BPTT method, the number of context neurons was set to $240$, corresponding to the total number of context neurons in the mixture of RNN experts model, and the parameters of the RNN were initialized in the same manner as for a module of the mixture of RNN experts model.

\begin{figure}
\begin{center}
\scalebox{1.0}{\includegraphics{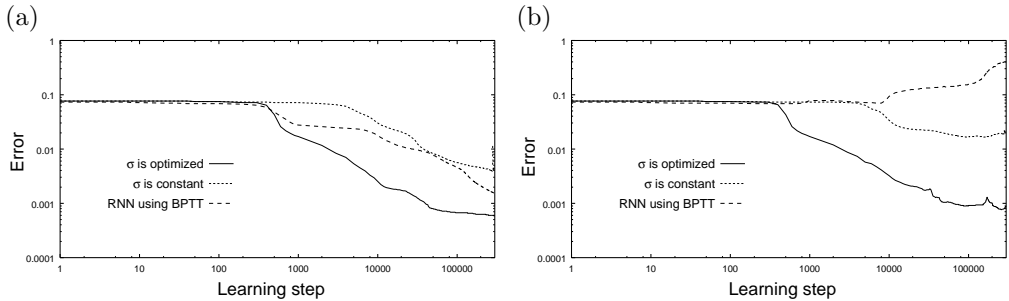}}
\caption{
Mean square error for each learning step.
(a) Open-loop dynamics.
(b) Closed-loop dynamics.}
\label{figure:error}
\end{center}
\end{figure}

Figure \ref{figure:error} displays the mean square error for each learning step, as defined by
\begin{equation}
E = \frac{1}{2Td} \sum_{n=1}^{T} ||\boldsymbol{y}_{n} - \boldsymbol{\bar{y}}_n||^2,
\end{equation}
where $\boldsymbol{y}_{n}$ and $\boldsymbol{\bar{y}}_n$ denote the training data and the output of the model, respectively.
The error of the trained model is evaluated by computing both the open-loop and closed-loop dynamics for each learning step.
The error for open-loop dynamics denotes the $\tau$-step prediction performance of the model.
As model learning reduces the error for open-loop dynamics by the equation (\ref{equation:gradient_vartheta}), this error represents the progress of learning.
However, from the point of view of temporal sequence learning, the error for closed-loop dynamics is a more important indicator of the performance in generating the desired sequences.
In general, even if the error for open-loop dynamics is sufficiently low, the error for closed-loop dynamics may not necessarily be reduced because temporal evolution may increase the margin of error via the self-feedback loop.
The reduction of error for closed-loop dynamics thus requires that the model successfully organize basins of attraction from the training data.

In Figure \ref{figure:error}, the error for open-loop dynamics for the RNN with BPTT is smaller than that for the mixture of RNN experts under constant $\boldsymbol{\sigma}$, whereas the error for closed-loop dynamics shows the opposite result.
This result implies that the RNN with BPTT is unable to construct basins of attraction corresponding to the training data, and so the RNN cannot generate sequences similar to the training data.
This result also demonstrates the reduction in the error for both open-loop and closed-loop dynamics by adaptively optimizing $\boldsymbol{\sigma}$.
The convergence of error in the proposed method was also achieved faster than in the conventional constant $\boldsymbol{\sigma}$ case.
The results indicate that the use of adaptively optimized $\boldsymbol{\sigma}$ in the learning scheme for the mixture of RNN experts model tends to provide better performance than achieved by the conventional and benchmark schemes from the viewpoint of sequence prediction and generation.
The results indicate that the mixture of RNN experts with adaptive $\boldsymbol{\sigma}$ tends to provide better performance than achieved by the conventional method and the standard RNN with BPTT method from the viewpoint of sequence prediction and generation.

\begin{figure}
\begin{center}
\scalebox{0.35}{\includegraphics{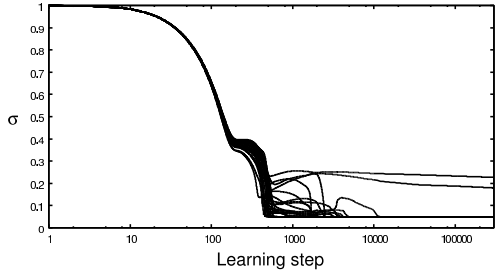}}
\caption{
The parameter $\boldsymbol{\sigma}$ under adaptive optimization.}
\label{figure:sigma}
\end{center}
\end{figure}

Figure \ref{figure:sigma} indicates that variances associated with many modules converged to the infimum $\bar{\sigma}$, because there was no additional noise in the training data.
Hence, it might be thought that the model is capable of learning the data without optimization of $\boldsymbol{\sigma}$, if $\sigma_i = \bar{\sigma}$ at the beginning.
Training under this condition, however, often fails.
The reason is that the parameter $\vartheta_i$ of each learning module $i$ diverges, because of the huge gradient $\Delta\theta(t)$.
These very large values for the gradient $\Delta\theta(t)$ occur because, in the initial learning phases, the mean square error is large but the variance is very small.

\begin{figure}
\begin{center}
\scalebox{0.35}{\includegraphics{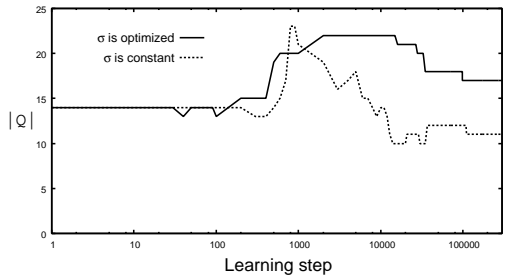}}
\caption{The number of elements in $Q$ for each learning step.}
\label{figure:classnum}
\end{center}
\end{figure}

The probability $q(i,n)$ that learning module $i$ is selected at time $n$ is 
\begin{equation}
q(i,n) = \frac{g_{n}^{(i)} p(\boldsymbol{x}_n | \gamma_i)}{\sum_{k=1}^{N} g_{n}^{(k)} p(\boldsymbol{x}_n | \gamma_k)} = \frac{g_{n}^{(i)} p(\boldsymbol{x}_n | \gamma_i)}{p(\boldsymbol{x}_n ~|~ \boldsymbol{\beta}_n, \boldsymbol{\gamma})}.
\end{equation}
Let us consider the function
\newcommand{\argmax}{\mathop{\rm argmax}\limits}
\begin{equation}
q_{\max}(n) = \argmax_{1 \leq i \leq N} q(i,n)
\end{equation}
that denotes the index of the module maximizing the probability $q(i,n)$, and a set $Q$ defined by
\begin{equation}
Q = \{i ~|~ \exists n ~ q_{\max}(n) = i \}.
\end{equation}
$|Q|$, the number of elements in the set $Q$, indicates the number of modules which can be effectively used to generate the sequence $Y$.
In Figure \ref{figure:classnum}, we have plotted $|Q|$ for each learning step.
There are two phases in the learning process. In the first phase $|Q|$ increases, and in the second phase it decreases.
This phenomenon often appears in the learning process when a mixture of RNN experts is utilized.
The reason for this behavior of $|Q|$ could be as follows:
In the first phase, while the mean square error is still large, the gradient $\Delta\theta(t)$ is directed to use learning modules to decrease the error; in the second phase, when the error has become small, the effect of the prior distribution becomes dominant and $|Q|$ decreases owing to suppression of the change of gate opening values.
Furthermore, in the case of adaptive $\boldsymbol{\sigma}$, $|Q|$ does not decrease until $\boldsymbol{\sigma}$ converges, because even if the error decreases, the variance $\boldsymbol{\sigma}$ also decreases by just that much, and therefore the values of the derivatives obtained by equations (\ref{equation:gradient_vartheta}) and (\ref{equation:gradient_sigma}) do not become small.
Thus, if $\boldsymbol{\sigma}$ is optimized, $|Q|$ decreases over the range for which the size of the error has been sufficiently reduced.
On the other hand, if $\boldsymbol{\sigma}$ is constant, then the error and  $|Q|$ decrease together.
However $|Q|$ can often decrease at a large rate, with a concomitant reduction in the rate at which the error decreases.

\begin{figure}
\begin{center}
\scalebox{1.0}{\includegraphics{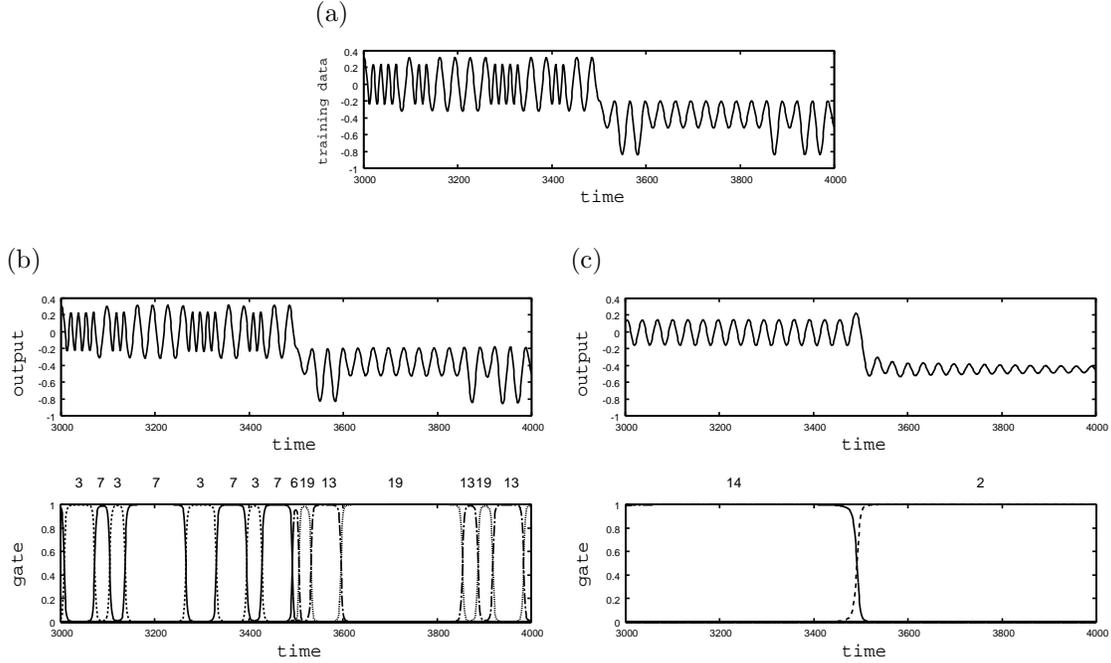}}
\caption{A snapshot of the training data, output and gate opening values at the end of learning.
(a) The training data.
(b) The case in which $\boldsymbol{\sigma}$ is optimized.
(c) The case of constant $\boldsymbol{\sigma}$.
In (b) and (c), the upper figures display output of trained models for the closed-loop dynamics, and lower figures display gate opening values, where the number over a gate opening value denotes the current opening gate.
}
\label{figure:gate}
\end{center}
\end{figure}

\begin{figure}
\begin{center}
\scalebox{1.0}{\includegraphics{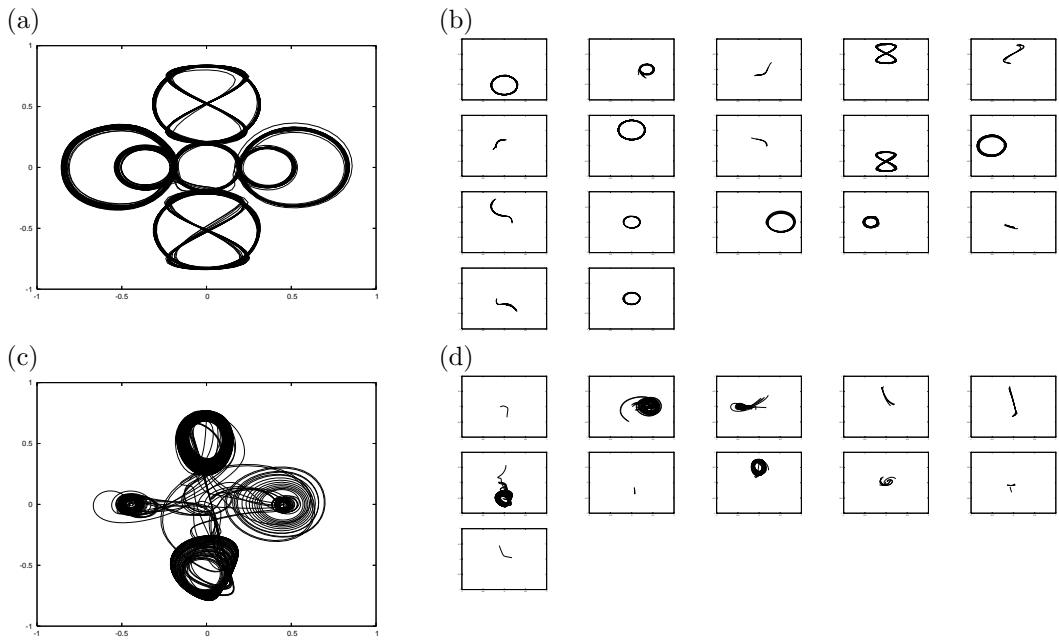}}
\caption{Trajectories generated by trained models in the closed-loop dynamics.
Here (a) and (b) display outputs of the trained model and the output of modules in the case of adaptive $\boldsymbol{\sigma}$, respectively.
(c) and (d) also display these outputs in the case of constant $\boldsymbol{\sigma}$.
Notice that the output of a module $i$ is plotted if $q_{\max}(n) = i$, namely, if gate $i$ opens at time $n$.
If gate $i$ never opened, then drawing the module $i$ is omitted.
}
\label{figure:output}
\end{center}
\end{figure}

Figure \ref{figure:gate} displays a snapshot of the training data, output and gate opening values at the end of learning.
Although the training data is a $2$-dimensional sequence, we have plotted only one dimension.
The result clearly shows that the trained model with adaptive $\boldsymbol{\sigma}$ can reconstruct spatio-temporal patterns in the training data successfully by adaptively switching the gates, but this is not so for the case with constant $\boldsymbol{\sigma}$.
Figure \ref{figure:output} (a) and (b) display the output of the trained model and the output of modules in the case of adaptive $\boldsymbol{\sigma}$, respectively.
Figures \ref{figure:output} (c) and (d) also display these outputs in the case of constant $\boldsymbol{\sigma}$.
It can be seen that the trained model successfully extracts the Lissajous curves as primitives in the adaptive $\boldsymbol{\sigma}$ case, whereas it could not extract the curves in the constant $\boldsymbol{\sigma}$ case..
These results indicate that the model with adaptive $\boldsymbol{\sigma}$ was able to segment the data into blocks corresponding to Lissajous curve patterns, and could allocate self-organized patterns in each corresponding module.
The model with constant $\boldsymbol{\sigma}$, however, was not able to perform this task effectively.

\subsection{Generalization} \label{subsection:generalization_error}

Here we examine the generalization capability of the proposed method as compared with that of the conventional method.
In order to evaluate this generalization capability, we prepared test data separate from the training data, and we regarded the mean square error for regenerating the test data as the generalization error.
However, the gate opening values determined by the parameter $\boldsymbol{\beta}_n$ correspond only to the training data.
Thus, in computing the generalization error, we reconstruct $\boldsymbol{\beta}_n$ that maximizes the likelihood for the test data by using the same learning scheme.
Note that the parameter $\vartheta_i$ of each learning module and the variance $\boldsymbol{\sigma}$ do not change in the reconstruction.
The aim of this analysis is to evaluate the general feasibility of the segmentation and data allocation scheme based on a gating mechanism.
If the gating mechanism successfully segments the training data into blocks as reusable primitives, and each module acquires a rule for allocating blocks in the training phase, not only the training error but also the generalization error will be small.
The value of $|Q|$ can also be regarded as a measure of generalization, representing the number of primitives extracted from the observed data.

To explain practical meaning of the generalization error and $|Q|$ more clearly, let us consider the case in which a gating network, which predicts or generates gate opening values, is added to the proposed model.
Although the definition of the model in section \ref{section:model} does not include the gating network because of focusing on the segmentation of temporal sequences, the gating network is usually used together with experts when applying to realistic problems.
In this case, if the generalization error of experts is sufficiently reduced, the generalization error of the whole system can be reduced by decreasing that of the gating network.
It is therefore sufficient to discuss the generalization ability of a single network when evaluating the generalization performance of the whole system, and the generalization ability of the single network can be evaluated by the stochastic approach \cite{Hammer2003}.
Hence, the model can be considered to have good generalization performance if it minimizes the generalization error.
Furthermore, training of the gating network tends to be easier when $|Q|$ is small.
Accordingly, the model can be understood to have good generalization performance if the scheme minimizes both the generalization error and $|Q|$.

This experiment serves to compare the case of adaptive $\boldsymbol{\sigma}$ with the constant case in which $\boldsymbol{\sigma} = \boldsymbol{1}$ and $\boldsymbol{\sigma} = \boldsymbol{\bar{\sigma}}$ ($= 0.05$).
Moreover, we can also compare them with the learning of RNN using BPTT, where the architecture of RNN is that of a mixture of RNN experts.
As training data and test data, we used a $2$-dimensional sequence generated by Markov chain switching of $2$ Lissajous curves of period $32$, with time delay of the feedback loop $\tau = 1$.
The lengths of the training and test data were $1000$ and $2000$, respectively.
In the optimizing $\boldsymbol{\sigma}$ case, we initialized $\boldsymbol{\sigma}$ to $5$.
Other parameter settings were the same as for Experiment 1.

\begin{figure}
\begin{center}
\scalebox{1.0}{\includegraphics{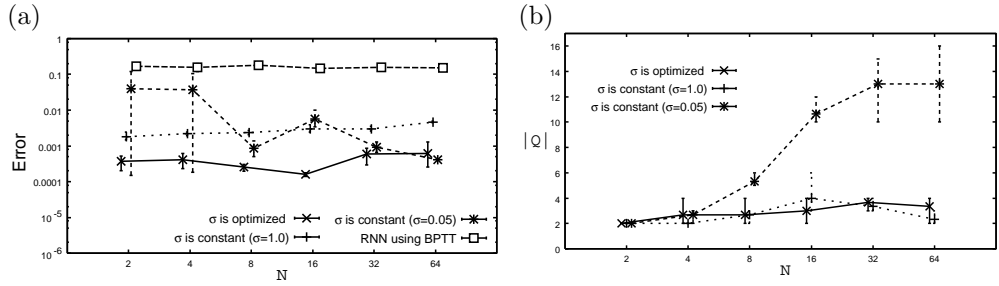}}
\caption{The generalization error and $|Q|$ after $100,000$ learning steps for each value of the parameter $N$, the number of learning modules.
(a) The generalization error for the closed-loop dynamics.
(b) The number of elements in the set $Q$.
In the case of RNN using BPTT, the number of context neurons in the RNN is set to $10N$, that is, the total number of context neurons in the mixture of RNN experts.
For each parameter $N$, we computed the results for $10$ samples with different initial conditions, training data and test data.}
\label{figure:error_and_classnum_for_each_N}
\end{center}
\end{figure}

The generalization error and $|Q|$ after $100,000$ learning steps are displayed in Figure \ref{figure:error_and_classnum_for_each_N}.
The results are shown for $10$ samples with different initial conditions for each $N$, the number of learning modules.
The generalization error for closed-loop dynamics is lower for the case of optimizing $\boldsymbol{\sigma}$ than for the other cases, and $|Q|$ for the optimizing $\boldsymbol{\sigma}$ is as low as that achieved by any of the other cases.
In the case of $\boldsymbol{\sigma} = \boldsymbol{\bar{\sigma}}$, even though the generalization error decreases with increasing $N$ to match that of the adaptive $\boldsymbol{\sigma}$ case, $|Q|$ increases markedly with $N$.
In the case of $\boldsymbol{\sigma} = \boldsymbol{1}$, neither the generalization error nor $|Q|$ are dependent on the number of modules, but the generalization error is greater than that in the adaptive $\boldsymbol{\sigma}$ case.
These results suggest that only the generalization error or $|Q|$, not both, can be reduced if $\boldsymbol{\sigma}$ is constant.
Thus, the model with constant $\boldsymbol{\sigma}$ is unable to continue performing effectively as the number of modules increases.
On the other hand, the adaptive optimization of $\boldsymbol{\sigma}$ satisfies the positive characteristics of both the $\boldsymbol{\sigma} = \boldsymbol{1}$ and $\boldsymbol{\sigma} = \boldsymbol{\bar{\sigma}}$ cases by reducing both generalization error and $|Q|$ even at large $N$.
For the reason which was explained in section \ref{subsection:learning_example}, by optimizing $\boldsymbol{\sigma}$ in the learning phase, $|Q|$ is low over the range in which the error margins are sufficiently reduced.
These results thus suggest that the proposed learning method for the mixture of RNN experts model is scalable with respect to the number of learning modules, maintaining good performance in terms of generalization error with increasing $N$.

\begin{figure}
\begin{center}
\scalebox{1.0}{\includegraphics{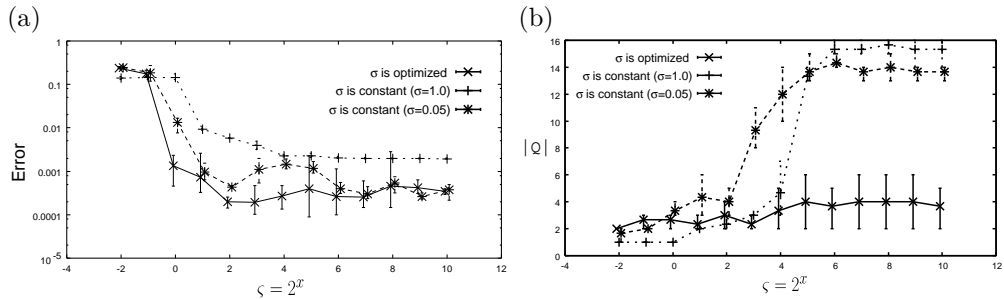}}
\caption{
(a)The generalization error for the closed-loop dynamics.
(b)The number of elements in the set $Q$ for the test data.
For each parameter $\varsigma$, we computed the results for $10$ samples up to $100,000$ learning steps, where the number of learning modules is $N = 16$.}
\label{figure:error_and_classnum_for_each_varsigma}
\end{center}
\end{figure}

Figure \ref{figure:error_and_classnum_for_each_varsigma} shows the effects of the parameter $\varsigma$ on generalization in learning.
The simulation computed up to $100,000$ learning steps for each parameter value $\varsigma$ of the prior distribution, where the number of modules $N$ was set at $16$.
The computation was repeated $10$ times for each $\varsigma$.
By equation (\ref{equation:gradient_beta}), the effect of suppressing gate opening change is inversely proportional to $\varsigma^2$.
If the change of gate opening is suppressed to an extreme degree, the model cannot learn the training data, and the generalization error is increased.
On the other hand, if the change of gate opening values is not suppressed, then $|Q|$ is not reduced.
From Figure \ref{figure:error_and_classnum_for_each_varsigma} (b), if $\boldsymbol{\sigma}$ is constant, dependence of $|Q|$ on the parameter $\varsigma$ is stronger than in the case of adaptive $\boldsymbol{\sigma}$.
As a result, in the case of optimizing $\boldsymbol{\sigma}$, the parameter range of $\varsigma$ for which both the generalization error and $|Q|$ are minimized, is larger than for the case of constant $\boldsymbol{\sigma}$.
Therefore, it can be said that the proposed method can achieve better generalization in learning more stably than the conventional one.

\subsection{Practical application} \label{subsection:humanoid_robot_experiment}

\begin{figure}
\begin{center}
\scalebox{0.75}{\includegraphics{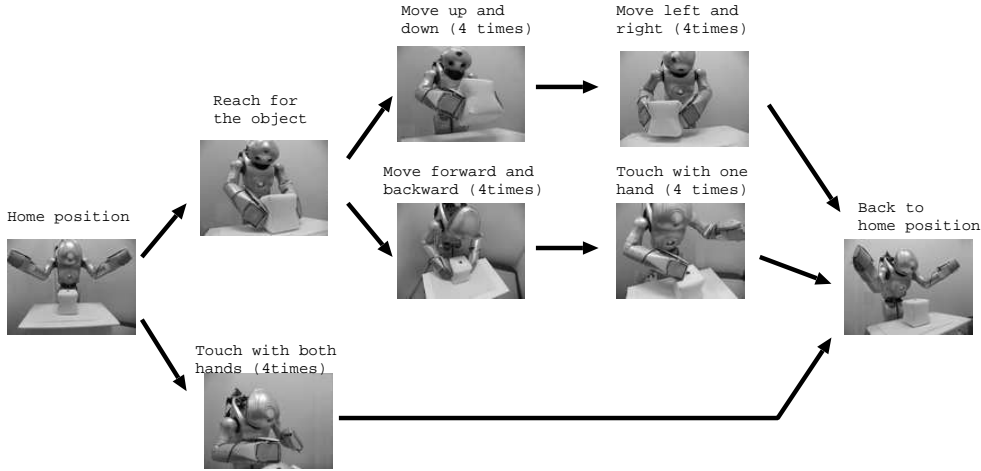}}
\caption{Humanoid robot behavior}
\label{figure:robot_motion}
\end{center}
\end{figure}

The practical learning performance of the proposed method is evaluated by applying the scheme to the learning of sensory-motor flows for a small humanoid robot.
The situation comprises a robot set before a workbench on which a cubic object was placed.
The task for the robot is to autonomously generate desired behaviours consisting of several primitive motions (see Figure \ref{figure:robot_motion}): (1) reaching for the object, (2) moving the object up and down, (3) moving the object left and right, (4) moving the object forward and backward, (5) touching the object with the left and right hand alternately, and (6) touching the object with both hands.
For each behaviour, the robot begins from a home position and ends with the home position.
Each behaviour is described by a $10$ dimensional sensory-motor flow, which consists of an $8$ dimensional motor vector representing the arm joint angle, and $2$ dimensional vision sense representing the object position.
The task is to predict and generate sensory-motor sequences.

In this section, a gating network to predict or generate gate opening values is added to the mixture of RNN experts model in order to realize a practically applicable scheme.
In the analyses above, gate opening values were determined by the parameter $\boldsymbol{\beta}_n$, because the main focus of this paper is the learning process to segment sequence data into primitives with respect to spatio-temporal patterns.
Since segmenting sequences is one of the most difficult processes in the learning of the mixture of experts model, our approach focusing on the data segmentation is efficient to investigate the learning processe.
However, such a model is unable to predict or generate unknown time series.
To apply the model to an actual problem, the model must be extended to a fully autonomous sequence generator by adding a gating network for module switching.
Thus, we consider the gating network learning in this section.
Learning for the gating network can be performed by adopting the gate opening vector given by $\boldsymbol{\beta}_n$ as a target.
The definitions of the gating network and the corresponding learning method are described in appendix \ref{section:appendix_B}, and further details can be found in a study \cite{Namikawa2008b}.

The training data consists of $3$ kinds of behaviors as depicted in Figure \ref{figure:robot_motion}, and $3$ initial locations of the object for each behavior, yielding a total of $9$ training sequences.
The model is constructed with $N = 16$ experts, $20$ context neurons for each expert, and $30$ context neurons for the gating network.
The time constant and time delay of the feedback loop are set to $\epsilon = \epsilon_g = 0.05$ and $\tau = 3$, respectively.
Other parameters are the same as in section \ref{subsection:learning_example}, and the parameters of the gating network are initialized in the same way as for a module of the mixture of RNN experts.

\begin{figure}
\begin{center}
\scalebox{1.0}{\includegraphics{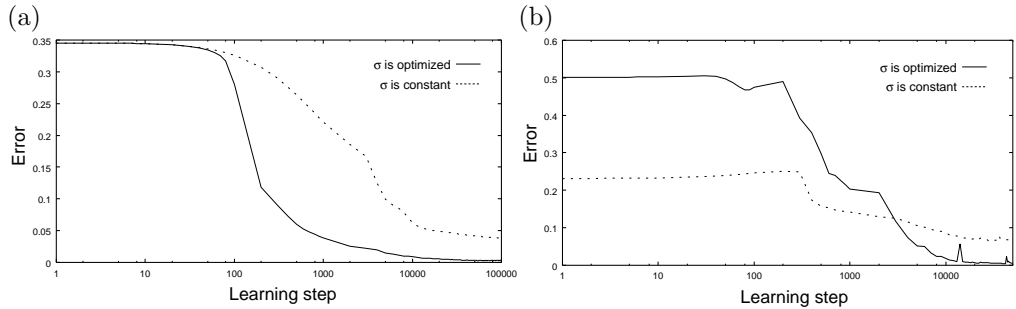}}
\caption{
Mean square error for closed-loop dynamics for learning of humanoid robot tasks.
(a) Learning for expert modules.
(b) Learning for a gating network to generate $\boldsymbol{g}_n$ in computation of closed-loop dynamics.
}
\label{figure:robot_error}
\end{center}
\end{figure}

\begin{figure}
\begin{center}
\scalebox{0.9}{\includegraphics{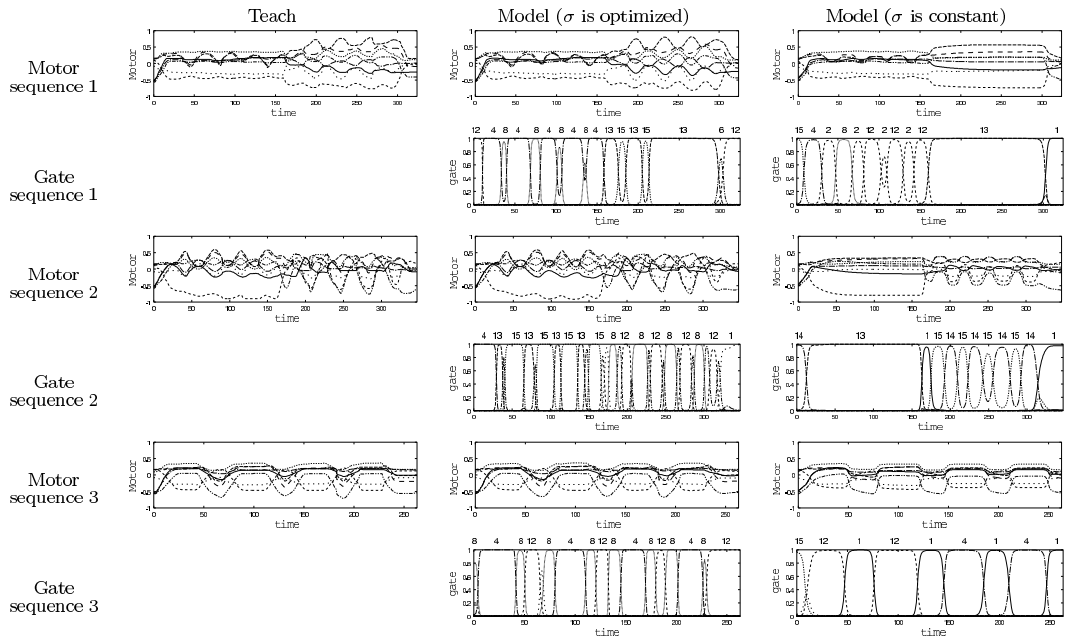}}
\caption{Time series of motor vector and gate opening vector.
Output $\boldsymbol{y}_n$ and gate opening vector $\boldsymbol{g}_{n}$ of trained model are computed in closed-loop dynamics.
For each time series, only the initial state of the model differs.}
\label{figure:robot_teach_and_output}
\end{center}
\end{figure}

Figure \ref{figure:robot_error} shows the mean square error for the closed-loop dynamics after training the mixture of RNN experts model up to $100,000$ steps and the gating network up to $50,000$ steps.
Figure \ref{figure:robot_teach_and_output} describes the training data, output, and gate opening values of the model trained with the gating network.
It can be seen that the model can autonomously generate sequences similar to the training data if $\boldsymbol{\sigma}$ is optimized by using our method, whereas the trained model is unable to generate similar sequences if $\boldsymbol{\sigma}$ is constant.
These results demonstrate that the proposed learning method is effective in application to realistic problems of time series prediction and generation, even when employing a gating network.

\section{Discussion} \label{section:discussion}

\subsection{Segmentation of temporal time series caused by indeterminacy} \label{subsection:segmentation_caused_by_indeterminacy}

One of the most important problems in the learning of a mixture of RNN experts model concerns the segmentation of sequence data into blocks as reusable primitives.
By the definition of the likelihood function, the module with smallest error margin is usually selected by the gating mechanism.
Moreover, as the prior distribution given by equation (\ref{equation:prior_distribution}) has the effect of suppressing the change in gate opening values, the gating mechanism retains the selected module until the module is no longer able to correctly predict sequence data.
Thus, locally represented knowledge of a reusable primitive is brought by the indeterminacy of the sequence data, and is embedded into each module.
Indeed, it was shown above that a mixture of RNN experts model is able to learn to generate data constructed by stochastically combining $9$ Lissajous curves.
Figure \ref{figure:output} (b) indicates that the adaptive optimization of $\boldsymbol{\sigma}$ allows a set of reusable chunks corresponding to the Lissajous curve patterns to be successfully extracted, and self-organized chunks to be allocated in each corresponding module.

The segmentation mechanism caused by the indeterminacy has also been discussed in a study by Tani and Nolfi \cite{Tani99}.
In their study, involving a robot actively exploring two rooms connected by a door, the sensory-motor flow was segmented by means of the uncertainty of the door opening.
Similarly, for stochastic switching between Lissajous curves or the arbitrary composition of robot motions in the present study, both of which involve indeterminacy in the observable data, the proposed learning method allows the mixture of RNN experts model to successfully segment the data using the information of nondeterministic switching.
The present scenarios thus reproduce the essential characteristics of the scenario considered by Tani and Nolfi in terms of the segmentation mechanism.

\subsection{Dynamic change of functions} \label{subsection:dynamic_change_of_functions}

As training data in section \ref{section:simulation}, we have considered Markov chain switching of Lissajous curves.
Switching patterns by Markov chain belongs to a simple class of systems in which a function dynamically changes.
In addition to such Markov chain models, there are other systems utilizing dynamic functions.
For instance, a switching map system is a dynamical system containing several maps, in which maps to govern temporal evolution of the system are dynamically switched with other maps in the system \cite{Sato2000,Namikawa2004}.
As other examples, we can consider chaotic itinerancy and function dynamics. Chaotic itinerancy is a class of phenomena such as chaotic itinerant motion among varieties of ordered states \cite{Ikeda89,Kaneko2003}, and a function dynamics \cite{Kataoka2000,Kataoka2001,Kataoka2003} is a dynamical system on a $1$-dimensional function space.
It is still unclear whether these systems can be learned by the model using the proposed method, either because the primitive sequences are governed by a chaotic dynamics, or the rule governing the change of primitives has a long-term dependency, more complex than that of a Markov chain.
Further analysis to explain learnability of the model is a future research topic.

\section{Conclusion} \label{section:conclusion}

In this study, we have presented a novel method of adaptive variance applied to learning for a mixture of RNN experts.
In order to show the capability of the proposed method, we have shown an example of a learning experiment in which the model using the proposed method can learn training data with respect to correctly segmenting a continuous flow into primitives, whereas the model using the conventional method cannot learn it.
The generalization capability has also been examined in order to compare the proposed method with the conventional method.
From these results, we claim that our method has improved learning performance of the model.
Furthermore, we have shown a humanoid robot experiment to make sure of a potential of our method.
This experiment has inferred that the model utilizing our method can be applied to realistic problems of time series prediction and generation.

\appendix
\section{Training data: Markov chain switching among nine Lissajous curves} \label{section:appendix_A}

The training data in section \ref{subsection:learning_example} comprise a $2$-dimensional sequence generated by Markov chain switching of $9$ Lissajous curves (see Figure \ref{figure:teaching_data}).
The equation for Lissajous curve $i$ is given by 
\begin{equation}
x_{n,1} = A_i\cos(a_i n) + B_i,
\end{equation}
\begin{equation}
x_{n,2} = C_i\sin(b_i n + c_i) + D_i,
\end{equation}
and the parameters $a_i, b_i, c_i, A_i, B_i, C_i, D_i$ are determined such that the period of each Lissajous curve is $32$.
The transition probability $R$ governing the change between Lissajous curves is defined by
\begin{equation}
R = {\small \left[ \begin{array}{ccccccccc}
\frac{1}{2} & \frac{1}{10} & \frac{1}{10} & \frac{1}{10} & \frac{1}{10} & \frac{1}{10} & \frac{1}{10} & \frac{1}{10} & \frac{1}{10} \\
\frac{1}{16} & \frac{9}{20} & \frac{9}{20} & 0 & 0 & 0 & 0 & 0 & 0 \\
\frac{1}{16} & \frac{9}{20} & \frac{9}{20} & 0 & 0 & 0 & 0 & 0 & 0 \\
\frac{1}{16} & 0 & 0 & \frac{9}{20} & \frac{9}{20} & 0 & 0 & 0 & 0 \\
\frac{1}{16} & 0 & 0 & \frac{9}{20} & \frac{9}{20} & 0 & 0 & 0 & 0 \\
\frac{1}{16} & 0 & 0 & 0 & 0 & \frac{9}{20} & \frac{9}{20} & 0 & 0 \\
\frac{1}{16} & 0 & 0 & 0 & 0 & \frac{9}{20} & \frac{9}{20} & 0 & 0 \\
\frac{1}{16} & 0 & 0 & 0 & 0 & 0 & 0 & \frac{9}{20} & \frac{9}{20} \\
\frac{1}{16} & 0 & 0 & 0 & 0 & 0 & 0 & \frac{9}{20} & \frac{9}{20} \\
\end{array} \right]},
\end{equation}
and it is assumed that the transition among curves is consonant with continuity of the orbit.

\section{Learning process of the gating network} \label{section:appendix_B}

We present the definition of a gating network together with the training procedure.
The dynamics of the gating network are defined by
\begin{equation} \label{equation:mixture_of_rnn_experts5}
\boldsymbol{u}_{n}^{g} = \big(1-\epsilon^{g}\big)\boldsymbol{u}_{n-1}^{g} + \epsilon^{g}\big( W_1^{g} \boldsymbol{g}_{n-\tau} + W_2^{g} \boldsymbol{x}_{n-\tau} + W_3^{g} \boldsymbol{c}_{n-1}^{g} + \boldsymbol{v}_1^{g}\big),
\end{equation}
\begin{equation} \label{equation:mixture_of_rnn_experts6}
\boldsymbol{c}_n^{g} = \tanh(\boldsymbol{u}_{n}^{g}),
\end{equation}
\begin{equation} \label{equation:mixture_of_rnn_experts7}
\boldsymbol{b}_n = W_4^{g} \boldsymbol{c}_{n}^{g} + \boldsymbol{v}_2^{g},
\end{equation}
\begin{equation} \label{equation:mixture_of_rnn_experts8}
g_{n}^{(i)} = \frac{\exp(b_n^{(i)})}{\sum_{k=1}^N \exp(b_n^{(k)})},
\end{equation}
where $\boldsymbol{x}_n$, $\boldsymbol{c}_n^{g}$ and $\boldsymbol{g}_n$ denote input states, context states and output states, respectively.
In addition, $\boldsymbol{u}_{n}^{g}$ and $\boldsymbol{b}_n$ are the internal potentials of neurons, and $\tau$ represents the time delay of feedback.
The output $\boldsymbol{g}_n$ of the gating network represents the gate opening vector at time $n$.
To discriminate between the gate opening vector determined by parameter $\boldsymbol{\beta}_n$ and the output of the gating network, in this appendix the gate opening vector determined by $\boldsymbol{\beta}_n$ is denoted as $\hat{\boldsymbol{g}}_n$, and is used as an estimate of the gate opening vector.

Let $\boldsymbol{\theta}^{g} = \big(W_1^{g}, W_2^{g}, W_3^{g}, W_4^{g}, \boldsymbol{v}_1^{g}, \boldsymbol{v}_2^{g}, \boldsymbol{u}_0^g\big)$ be a set of learnable parameters of a gating network.
Assume that $D^{g} = (X, (\hat{\boldsymbol{g}}_n)_{n=1}^{T})$ is given.
We define a likelihood $L(D^{g}, \boldsymbol{\theta}^{g})$ by the Dirichlet distribution
\begin{equation}
p(\boldsymbol{g}_n ~|~ \hat{\boldsymbol{g}}_n) = \frac{1}{Z(\hat{\boldsymbol{g}}_n)}\prod_{i=1}^{N}\big(g_n^{(i)}\big)^{\hat{g}_n^{(i)}},
\end{equation}
\begin{equation} \label{equation:likelihood_of_gating_net}
L(D^{g}, \boldsymbol{\theta}^{g}) = \prod_{n=1}^{T} p(\boldsymbol{g}_n ~|~ \hat{\boldsymbol{g}}_n),
\end{equation}
where $Z(\hat{\boldsymbol{g}}_n)$ is the normalization constant, and $\boldsymbol{g}_n$ is given by $X$ and $\boldsymbol{\theta}^{g}$.
Notice that $\boldsymbol{g}_n$ in equation (\ref{equation:likelihood_of_gating_net}) is calculated by the teacher forcing technique \cite{Zipser89}, namely, using $\hat{\boldsymbol{g}}_{n-\tau}$ instead of $\boldsymbol{g}_{n-\tau}$ in equation (\ref{equation:mixture_of_rnn_experts5}).
In addition, because of the term $\boldsymbol{x}_{n-\tau}$ on the right-hand side of equation (\ref{equation:mixture_of_rnn_experts5}), $\boldsymbol{g}_n$ cannot be computed if $n \leq \tau$. 
It is therefore assumed that $\boldsymbol{g}_n = \hat{\boldsymbol{g}}_n$ if $n \leq \tau$.
If $g_n^{(i)}$ is considered to be the probability of choosing an expert $i$ at time $n$, maximizing $\ln L(D^{g}, \boldsymbol{\theta}^{g})$ is equivalent to minimizing the Kullback-Leibler divergence
\begin{equation}
D_{\mathrm{KL}}((\hat{\boldsymbol{g}}_n)_{n=1}^{T} || (\boldsymbol{g}_n)_{n=1}^T) = \sum_{n=1}^{T}\sum_{i=1}^{N} \hat{g}_n^{(i)} \ln\Big(\frac{\hat{g}_n^{(i)}}{g_n^{(i)}}\Big).
\end{equation}

The learning process for the gating network involves to choose the best parameter $\boldsymbol{\theta}^{g}$ by using the gradient descent method for the likelihood $L(D^{g}, \boldsymbol{\theta}^{g})$, in the same way as the equations (\ref{equation:gradient_descent_method_with_momentum1}) and (\ref{equation:gradient_descent_method_with_momentum2}).
The partial differential equation $\frac{\partial \ln L(D^{g}, \boldsymbol{\theta}^{g})}{\partial \boldsymbol{\theta}^{g}}$ is given by
\begin{equation}
\frac{\partial \ln L(D^{g}, \boldsymbol{\theta}^{g})}{\partial \boldsymbol{\theta}^{g}} = \sum_{n=1}^{T} \sum_{i=1}^{N}\frac{\hat{g}_n^{(i)}}{g_n^{(i)}} \big( \frac{\partial g_n^{(i)}}{\partial \boldsymbol{\theta}^{g}} \big).
\end{equation}
Note that $\frac{\partial g_n^{(i)}}{\partial \boldsymbol{\theta}^{g}}$ can be solved by the BPTT method.
The procedure for training the model consisting of the mixture of RNN experts and gating network is organized into two phases as follows:
\begin{ol}{2}
\item We compute the experts learning together with an estimate $\hat{\boldsymbol{g}}_n$ of the gate opening vector.
\item Using the estimate $\hat{\boldsymbol{g}}_n$ of gate opening vector as a target, we compute learning for the gating network.
\end{ol}
The training procedure progresses with phase (2) after convergence of phase (1).

\bibliography{bibliography07}

\end{document}